\documentclass[aps,prc,twocolumn,fleqn,showpacs,showkeys]{revtex4}

\usepackage{amsmath, amssymb}
\usepackage{graphicx,color}

\begin{document}

\title{Jet-Triggered Back-Scattering Photons for Quark Gluon Plasma Tomography}

\author{Somnath De$^{1,2}$}
\author{Rainer J.\ Fries$^{1}$}
\author{Dinesh K.\ Srivastava$^{2}$}

\affiliation{$^1$Cyclotron Institute and Department of Physics and Astronomy,
  Texas A\&M University, College Station, TX 77845, USA}
\affiliation{$^2$Variable Energy Cyclotron Center, Kolkata, India}


\begin{abstract}
High energy photons created from back-scattering of jets in
quark gluon plasma are a valuable probe of the temperature of the plasma, 
and of the energy loss mechanism of quarks in the plasma. An unambiguous 
identification of these photons through single inclusive photon measurements 
and photon azimuthal anisotropies has so far been elusive. We estimate the 
spectra of back-scattering photons in coincidence with trigger jets for 
typical kinematic situations at the Large Hadron Collider and the 
Relativistic Heavy Ion Collider. We find that the separation of 
back-scattering photons from other photon sources using trigger 
jets depends crucially on our ability to reliably estimate the
initial trigger jet energy. We estimate that jet reconstruction techniques
in heavy ion experiments need to be able to get to jet $R_{AA}\gtrsim 0.7$
in central collisions for viable back-scattering signals.
\end{abstract}


\maketitle

\section{Introduction}

Electromagnetic radiation has a long history as an excellent probe in high
energy nuclear collisions. The long mean free path of  photons and dileptons,
an order of magnitude larger than the transverse size of the colliding nuclei,
allows them to carry information from the earliest stages of the collision and 
from deep inside the fireball to the detector systems. Over the years, several 
distinct sources of direct photons have been identified and calculated. They 
include (i) prompt photons from initial hard processes between beam partons
and from jet fragmentation \cite{Owens:1986mp,Catani:2002ny,Aurenche:2006vj}, 
(ii) pre-equilibrium photons
from the secondary scatterings between partons before the system thermalizes 
\cite{Bass:2002pm}, (iii) photons from jets interacting with QGP 
\cite{Fries:2002kt,Fries:2005zh,Zakharov:2004bi},  (iv) thermal radiation from 
equilibrated or near-equilibrium quark gluon plasma (QGP) 
\cite{Kapusta:1991qp,Baier:1991em,Aurenche:1998nw,Arnold:2001ms}, 
(v) photons associated with the hadronization process \cite{ChenFries:2013}, 
and finally (vi) thermal photons from the hot hadronic gas phase 
\cite{Kapusta:1991qp,Turbide:2003si}. These direct photons have to be 
experimentally separated from a large amount of background photons from 
hadronic decays (most notably from neutral pions).

Thermal photons, dominant at low transverse momenta $p_T$, are supposed to 
act as a thermometer of the hot nuclear matter, and there is mounting evidence 
that the early temperatures extracted are above the critical temperature $T_c$ 
expected for the phase transition to quark gluon plasma
\cite{Adare:2008ab,Wilde:2012wc}. Photons from interactions of jets with QGP 
carry important complementary information. Hence it is critical to 
experimentally separate the contributions from different photon 
sources as much as possible so each can be analyzed appropriately. 
The list of photon sources in the previous paragraph follows a rough hierarchy
of typical transverse momenta of the source, from high to low $p_T$.
Jet-medium photons have been shown to make significant contributions at
intermediate $p_T$ around $\sim 4$ GeV/$c$ in single inclusive photon spectra
both at the Relativistic Heavy Ion Collider (RHIC) and the Large Hadron
Collider (LHC), but they compete with prompt hard photons at larger $p_T$ and 
thermal and pre-equilibrium photons at smaller $p_T$. Hence it has been hard 
to confirm their existence from measurements of single inclusive photon 
spectra alone, much less to exploit their properties. Elliptic flow of 
jet-medium photons had been predicted to be negative and it was expected to 
serve as a telltale signature \cite{Turbide:2005bz,Turbide:2007mi}. 
However experimental studies of direct photon $v_2$ have not been able to 
bring conclusive evidence for the existence of jet-medium photons 
\cite{Adare:2011zr,Lohner:2012ct}.

In this work we propose to use the correlation of large momentum photons
with jets in the opposite direction to measure the strength of a part of 
the jet-medium photon source, more precisely the photons from back-scattering 
kinematics.
We will argue that this effectively rids the sample of photons from thermal
and pre-equilibrium sources and vastly reduces the background from jet 
fragmentation photons. Furthermore, energy loss of the parent parton should 
shift back-scattering photons toward smaller momenta, exposing them 
compared to the remaining background source of prompt hard photons
which are not affected  by parton energy loss. On the other hand, energy loss
of the trigger jet, and the experimental uncertainty measuring jet energies
tend to wash out the signal from back-scattering photons.
We will discuss these effects in detail below.
Great opportunity awaits us if we successfully measure the strength of 
the back-scattering process. Besides having a complementary measure of 
parton energy loss independent of hadronic measurements (quarks will lose
energy before converting into photons), one could measure the temperature 
of the medium ($T\sim 200$ MeV) independently using back-scatter photons 
with energies of tens of GeV.

Jet-medium photons have most notably been calculated
in two limits: as electromagnetic bremsstrahlung to jet quenching
\cite{Zakharov:2004bi,Arnold:2001ms}, e.g.\ in the
Arnold-Moore and Yaffe (AMY) approach, and as an elastic back
scattering process \cite{Fries:2002kt}. The latter is based on the fact that 
$2 \to 2$
Compton and annihilation scattering with a photon in the final state,
$q+g \to \gamma+ q$ and $q+\bar q\to \gamma+g$ both have a sharp peak at 
backward angles. In other words, when a fast quark annihilates with a
slow antiquark, or Compton scatters off a slow gluon from the thermal medium,
in most of the cases (Compton) or in about half the cases (annihilation) 
the photon created carries approximately the momentum of the fast quark.
Back-scattering processes of this type are well known and are exploited
in numerous ways, e.g.\ to create high energy photon beams. In photon beam 
facilities laser photons (typically $\sim 1$ eV) are Compton back-scattered 
from a high energy electron beam (in the MeV to GeV range) to create 
collimated beams of MeV to GeV photons 
\cite{Milburn:1962jv,Arutyuninan:1963aa}. The QCD Compton analogue that 
we use here consists of a thermal gluon ($\sim 200$ MeV) scattering 
off a quark ($\sim 20$ GeV) to produce a $\sim 20$ GeV photon.
Both bremsstrahlung and back-scattering calculations are often carried out
in a leading parton approximation to jets in a medium \cite{Fries:2002kt}
however more general calculations using the parton dynamics inside a jet shower
in a medium have recently become available \cite{Renk:2013kya}.

\section{Calculating Photon Sources}

In Ref.\ \cite{Fries:2002kt} the rate of Compton and annihilation processes
between one parton from a set of fast quarks subject to energy loss, and 
another from a fireball with a temperature profile 
$T(x) = T(\tau,\eta,\mathbf{x}_\perp)$ was 
calculated in the backward peak approximation ($\mathbf{p}_\gamma \approx
\mathbf{p}_{\mathrm{fast}\, q}$) to be 
\begin{multline}
 \label{eq:master}
  E_\gamma \frac{dN}{d^4x d^3p_\gamma} = \frac{\alpha\alpha_s}{4\pi^2}
  \sum_{q=1}^{N_f} \left(\frac{e_q}{e}\right)^2  T^2(x)\\ \times
  \left[ f_q(\mathbf{p}_\gamma,x) + f_{\bar q}(\mathbf{p}_\gamma,x) \right]
  \left[ \ln \frac{3E_\gamma }{\alpha_s\pi T(x)} + C\right],
\end{multline}
where $C=-1.916$. Here $\alpha$ and $\alpha_s$ are the electromagnetic and 
strong coupling constant respectively. $f_q$ is the phase space distribution 
of fast quarks interacting with the medium and $e_q$ is the electric charge 
of a quark with the index $q$ running over all active quark flavors. 
This formula is easily generalized to the rate of photons associated with a 
trigger jet whose energy, pseudorapidity and relative azimuthal angle, 
$E_T$, $y_j$, $\phi_j$,
fall within a trigger window $\mathcal{T}_j$ in $E_T$-$y_j$-$\phi_j$ space.
For the latter we replace the single inclusive parton distribution
$f_{q}(\mathbf{p}_\gamma,x)$ by the parton-jet pair distribution integrated over
$\mathcal{T}_j$,
\begin{multline}
  f^{\mathcal{T}_j}_{q}(\mathbf{p}_q,x) = \frac{(2\pi)^3}{g_q\tau p_{T}} 
  \delta(y-\eta) \rho(\tau,\mathbf{x}_{\perp}^0)       \\ \times
  \int_{\mathcal{T}_j} dE_Tdy_jd\phi_j   E_q \frac{dN}{d^3p_q
  dE_T dy_{j}d\phi_{j}}\Big|_{{\mathbf{p}_q^0=\mathbf{p}_q+\Delta
      \mathbf{p}_q}\atop{E_T^0=E_T + \Delta E_T}} \, .
\end{multline}
Here $x=(\tau,\eta,\mathbf{x}_\perp)$ and $\mathbf{p}_q$ are the position
and momentum of the quark at the time of the back scattering and
$x^0 = (\tau_0,\eta,\mathbf{x}_\perp^0)$ and $\mathbf{p}_q^0$ are the 
original position and momentum when the quark was created in a hard process.
Propagation is assumed to be along straight lines in the direction of 
$\mathbf{p}_q$ with the speed of light, i.e. 
$\mathbf{x}_\perp = \mathbf{x}_\perp^0 + (\tau-\tau_0) \mathbf{\hat p}_q^0$. 
$\Delta \mathbf{p}_q =\mathbf{p}_q^0 - \mathbf{p}_q$ is the energy lost 
between $\mathbf{x}_\perp^0$ and $\mathbf{x}_\perp$. Straight line propagation
implies that $\Delta \mathbf{p}_q$ is collinear with the original 
momentum of the quark. Similarly $\Delta E_T$ is the energy lost by the
trigger jet in the medium. $\Delta E_T$ will strongly depend on the cone size
chosen in the experimental reconstruction of the jet.
$g_q=6$ is the spin and color degeneracy factor of quarks and $\rho$ is the 
density of nucleon-nucleon collisions in the transverse plane.

We have to consider the background from prompt hard photons and fragmentation 
photons with an awayside jet in the same trigger window $\mathcal{T}_j$. We
will not consider trigger windows with jet $E_T$ smaller than 20 GeV.
The pre-equilibrium and thermal photons do not possess back-to back correlation
with an away-side jet, hence can be eliminated from the background.
We can thus compute the nuclear modification factor $R_{AA}$ of photons
with an awayside high energy trigger jet as:
\begin{equation}
  \label{eq:raa}
  R_{AA} = \frac{\text{(backscat. + prompt hard +
      fragment.)}_{A+A}}{N_{\mathrm{coll}}\times\text{(prompt hard + fragment.)}_{p+p}}
  \, 
\end{equation}
where $N_{\mathrm{coll}}$ as usual is the total number of binary nucleon-nucleon
collisions.

Our calculation comprises two stages. At the first stage we calculate
the background (prompt hard and fragmentation) photon and parton 
(prior to back-scattering) cross sections
at leading order (LO) or next-to-leading
order (NLO) in $\alpha_s$ in the code JETPHOX (version 1.2.2) 
\cite{Catani:2002ny,Aurenche:2006vj}. The default will be LO 
cross sections unless explicitly stated otherwise. We use CTEQ6M 
\cite{Pumplin:2002vw} parton distributions for protons and EPS09 
modifications for nuclei \cite{Eskola:2009uj} in JETPHOX.

For the second stage we use the code package PPM 
\cite{Rodriguez:2010di,Fries:2010jd} to calculate (i) the energy loss of 
partons, (ii) the energy loss of jets, and (iii) the back-scattering photon
rate according to Eq.\ (\ref{eq:master}). PPM propagates partons and jets
(represented by their leading parton) through a fireball model. Here this is
done pairwise, i.e.\ photon-jet pairs and quark-jet pairs propagate from
their point of creation though a hard parton-parton scattering. 
The spatial distribution $\rho$ of hard processes is given by the 
nucleon-nucleon collision density from a Glauber calculation.
For photon-jet
pairs the energy loss of the jet due to its path through the medium is 
calculated, and all photon-jet pairs with a final jet energy within 
$\mathcal{T}_j$ are counted as part of the background. We do not take 
into account energy loss of partons before fragmentation into photons 
which will lead to a lower bound for the signal to background ratio. If 
energy loss of partons for photon fragmentation
were taken into account in addition, it would help to suppress the 
fragmentation background at high photon-$z$ where $z$ is the momentum 
fraction of the parent parton carried by the photon.
For quark-jet pairs the energy loss of the jet and of the 
parton are computed while the back-scattering probability of the parton 
is also computed along the way. All final photons from this source which
lie in $\mathcal{T}_j$ are counted as part of the photon signal.

Our fireball model describes a longitudinally expanding,
boost-invariant QGP phase. The transverse profile of the entropy density
is fixed by the participant density of nucleons from a Glauber calculation. 
We do not expect our main conclusions to change much if transverse expansion or 
fluctuations in the fireball are taken into account. The normalization of the 
entropy density is fixed by data from RHIC \cite{Pal:2003rz} and scaled up 
to describe multiplicity data in Pb+Pb collisions at the LHC. We use a 
relativistic ideal gas equation of state for 3 light quark flavors to 
calculate the temperature needed in the photon conversion formula. This 
procedure will slightly underestimates the real temperature and thus the
photon production rate at a given value of the entropy density $s$, in 
particular close to the pseudo-critical temperature $T_c$.

The energy loss of quarks and gluons is calculated 
from a simple LPM-inspired approximation (called sLPM in Ref.\ 
\cite{Rodriguez:2010di}) which uses $dp_T/d\tau = - \hat q (\tau-\tau_0)$ 
where the value of $\hat q$ is proportional to the local entropy density 
$s$ of the fireball at that space-time point. The proportionality constant 
is fitted to simultaneously describe RHIC and LHC data on single
inclusive hadron suppression. Despite its simplicity, this model describes
basic features of high momentum hadron production at RHIC reasonably well
\cite{Rodriguez:2010di}. The resulting initial value of 
$\hat q \approx 1.2$ GeV$^2$/fm in the center of Au+Au collisions at RHIC 
energy is consistent with recent findings of the JET collaboration 
\cite{Burke:2013yra}.

Jet energy loss is much less under theoretical control. A consistent 
calculation can only be done with a full jet shower simulation in the medium, 
e.g.\ \cite{Renk:2013kya}. Here we choose a simple model of the path length and
energy dependence to reproduce gross features of jet energy loss. 
We parameterize that the energy loss 
(i.e.\ the amount of energy outside of a given jet cone) 
is proportional to path length, and we add a small energy dependence, 
$dE_T/d\tau = - \hat r \ln(E_T/\Lambda)$ where $\Lambda = 0.2$ GeV.
$\hat r$ is proportional to the local entropy density $s$ as in the case
of leading parton energy loss.The linear path length dependence
appears more appropriate both for the stochastic process of stripping
partons off the jet cone as the jet goes through the medium, and for 
the large angle radiation with short formation times that plays a role as 
well. The normalization of $\hat r$ is varied  to obtain different 
inclusive jet $R_{AA}$.

\section{Results}

\begin{figure}[tb]
\begin{center}
\includegraphics[width=\columnwidth, clip=true]{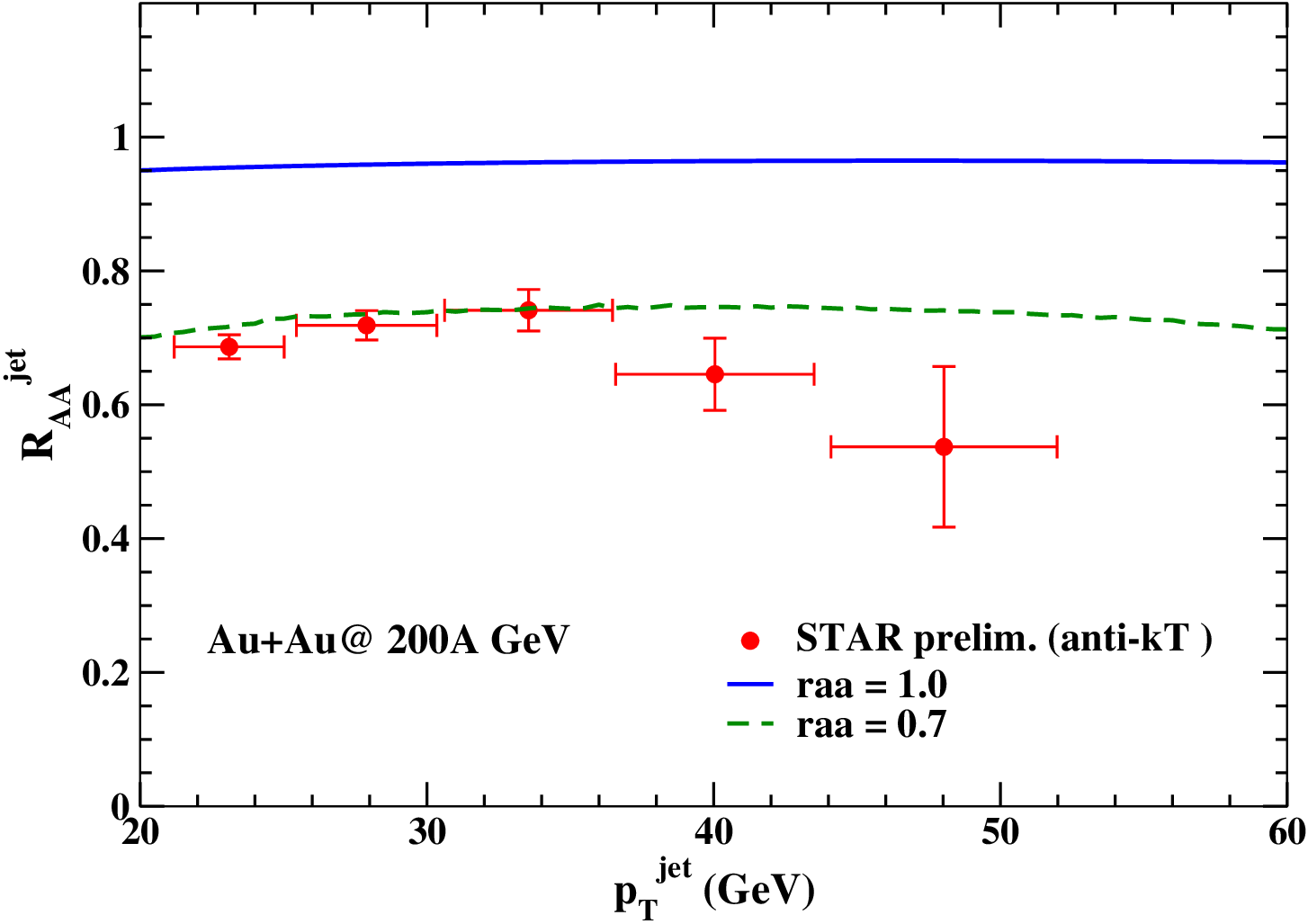}\\[2em]
\includegraphics[width=\columnwidth, clip=true]{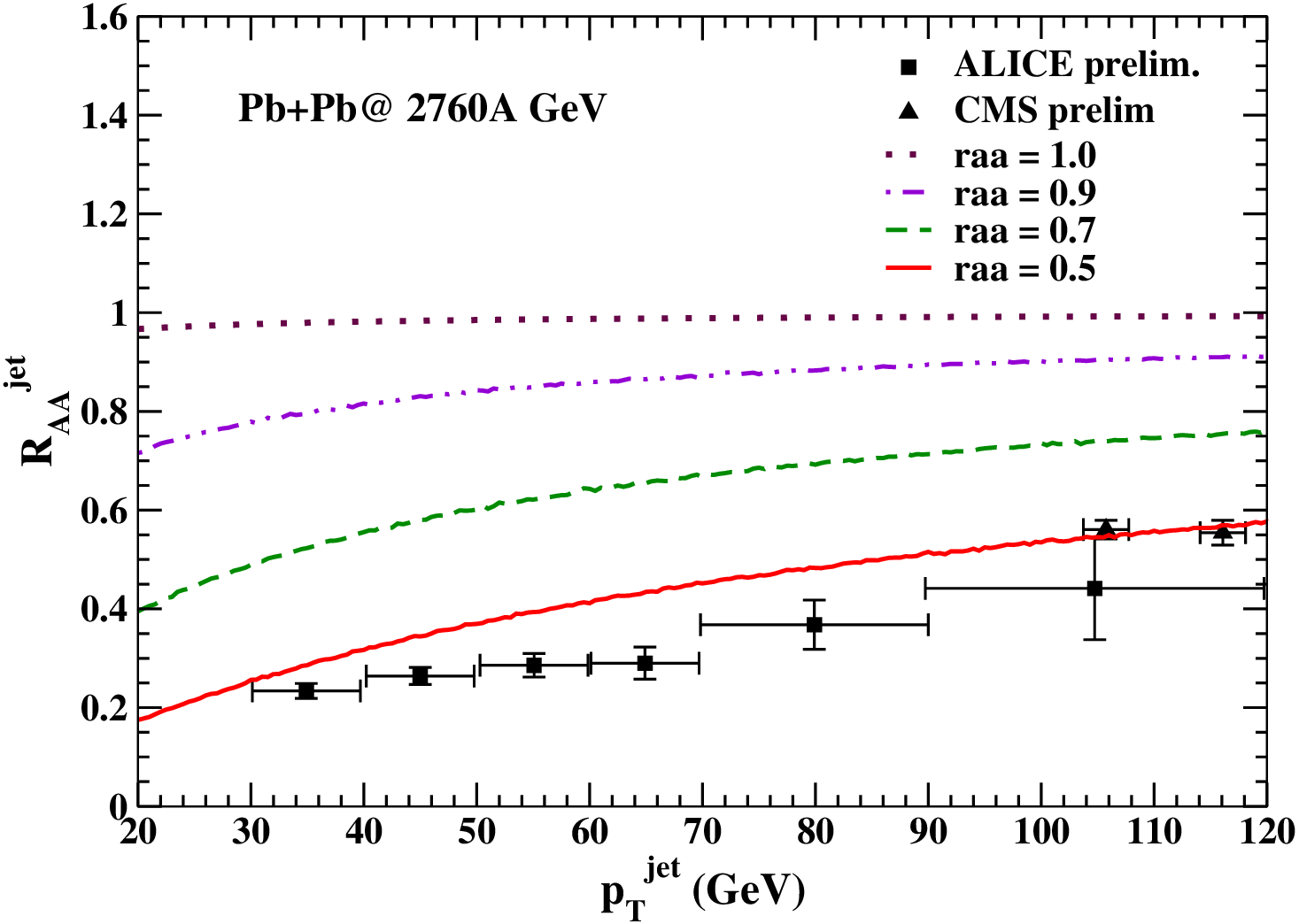}
\caption{(Color online) Upper Panel: $R_{AA}$ of single inclusive jets as a function of jet $p_T$ 
in central Au+Au collisions at RHIC for two values of $\hat r$ corresponding 
to ``raa'' values of roughly 1.0 ($\hat r =0$) and 0.7 at 30 GeV respectively.
Lower Panel: The same for central collisions of lead nuclei at LHC energy, for four 
values of $\hat r$, corresponding to raa values of roughly 1.0, 0.9, 0.7 
and 0.5 at 100 GeV respectively. Data from the STAR~\cite{Ploskon:2009zd},
ALICE~\cite{Reed:2013rpa} and CMS~\cite{BeltTonjes:2013fta} collaborations for
jet cone radii of 0.4, 0.2 and 0.4 respectively, are also shown for comparison.}
\label{fig:jetraa}
\end{center}
\end{figure}

In order to calibrate jet energy loss we calculate the nuclear modification 
factor $R_{AA}$ of single inclusive jets for both central Au+Au collisions at 
RHIC energy and central Pb+Pb collisions at LHC energy in our jet energy loss
model. This allows us to scale the normalization of the parameter $\hat r$ 
to reproduce a certain inclusive jet $R_{AA}$. We will refer to different
values of jet energy loss by quoting the approximate value of $R_{AA}$ at 
$E_T = 30$ GeV for RHIC and $E_T = 100$ GeV at LHC, respectively. We will
quote this number in plots as ``raa''.
Fig.\ \ref{fig:jetraa} shows the single inclusive $R_{AA}$ for jets 
corresponding to values of raa of roughly 1, 0.9, 0.7 and 0.5 for central 
Pb+Pb collisions at LHC and 1.0 and 0.7 for central Au+Au collisions at RHIC, respectively. 
We also show the data from STAR, ALICE and CMS that use rather small jet cone 
radii of 0.4, 0.2 and 0.4, respectively. 
Without a full jet shower simulation we can not make 
a rigorous connection between jet cone radius, jet quenching, and jet $R_{AA}$.
Rather we will present our results using a set of different values of 
``raa''
(and thus $\hat r$). As can be seen from the figure the lowest values of raa 
for both RHIC (0.7) and LHC (0.5) roughly correspond to the suppression seen
in current data with small cone radii. With improving jet 
reconstruction techniques and larger jet cone radii larger values of ``raa''
might become feasible.
Small jet cones in heavy ion experiments are mostly dictated by the 
relatively large background that needs 
to be subtracted. 
The value of $\hat r$ needed to reproduce raa of 0.7 at RHIC is 
about $0.24$ GeV/fm initially in the center of head-on Au+Au collisions 
which corresponds to an initial energy loss of $\sim 1.2$ GeV/fm for 30 GeV 
jets.

We can now proceed to calculate photon spectra opposite 
of trigger jets in several scenarios. For this preliminary study 
we choose the trigger window $\mathcal{T}_j$ for the jet to be defined as 
$-1 < y_j < 1$ and 30--35 GeV in $E_T$ for RHIC, and $-2 < y_j < 2$
and 60--65 GeV in $E_T$ for LHC. We define the away-side as an 
angle between 165 and 195 degrees in relative azimuthal angle. Let us briefly 
discuss the choice of trigger window. The yield of single-inclusive 
back-scattering photons falls faster with $p_T$ than
prompt hard photons (similar to a higher twist contribution in perturbative
QCD), thus the signal will become stronger with smaller 
$p_T$. However, experiments need to be able to reconstruct jets in a 
reliable manner. This puts a lower bound on the trigger window $E_T$. Our
choice is an attempt to maximize the back-scattering yield while keeping
jet reconstruction feasible. We would also like to make our back-scattering
signals as sharply defined as possible, which is ideally achieved with 
very narrow trigger windows. However, uncertainties in the jet energy 
reconstruction put constraints on the energy resolution achieved in 
experiments. We have chosen a trigger window width of 5 GeV for this study.

\begin{figure}[tb]
\begin{center}
\includegraphics[width=\columnwidth, clip=true  
]{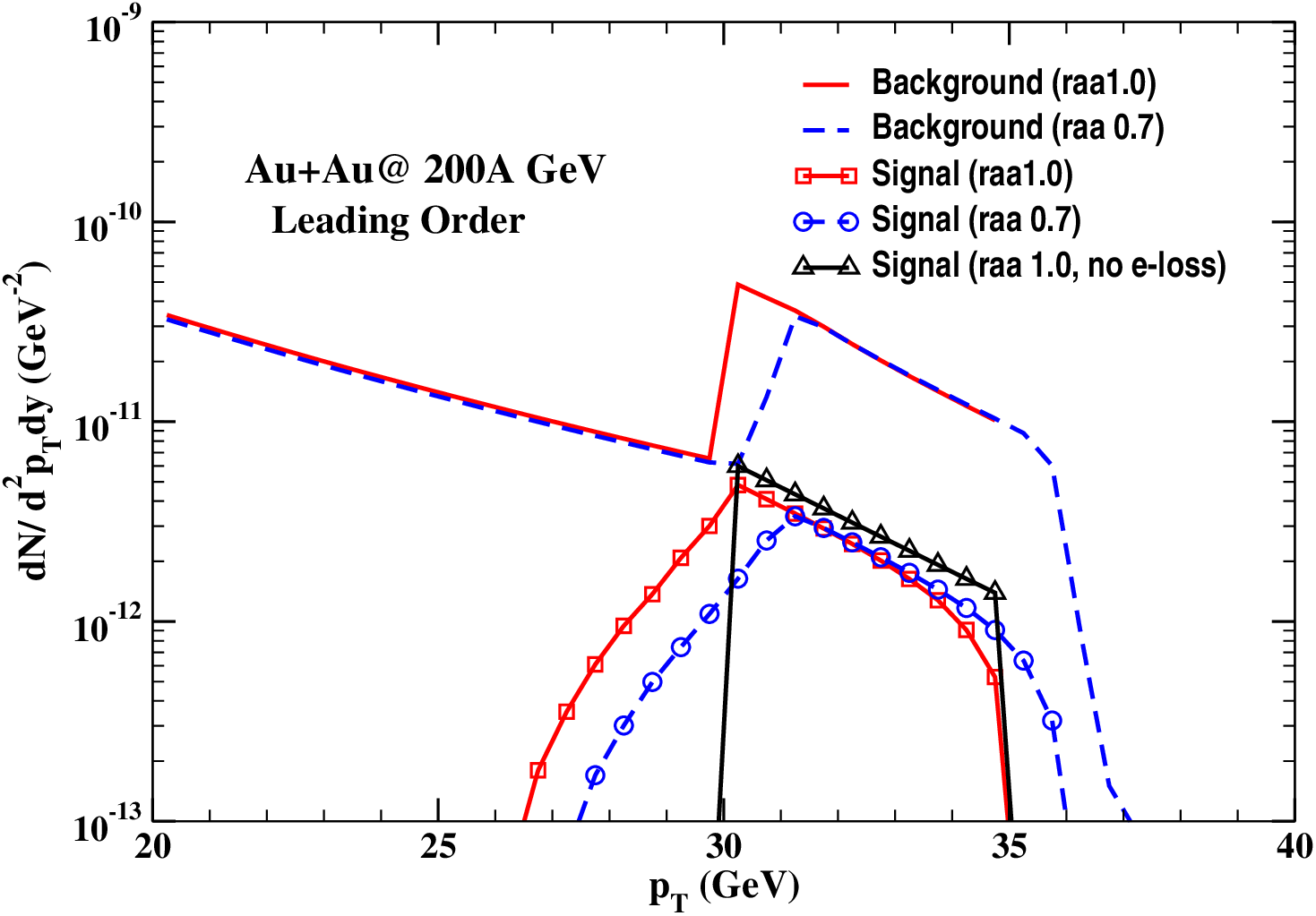}\\[2em]
\includegraphics[width=\columnwidth, clip=true]{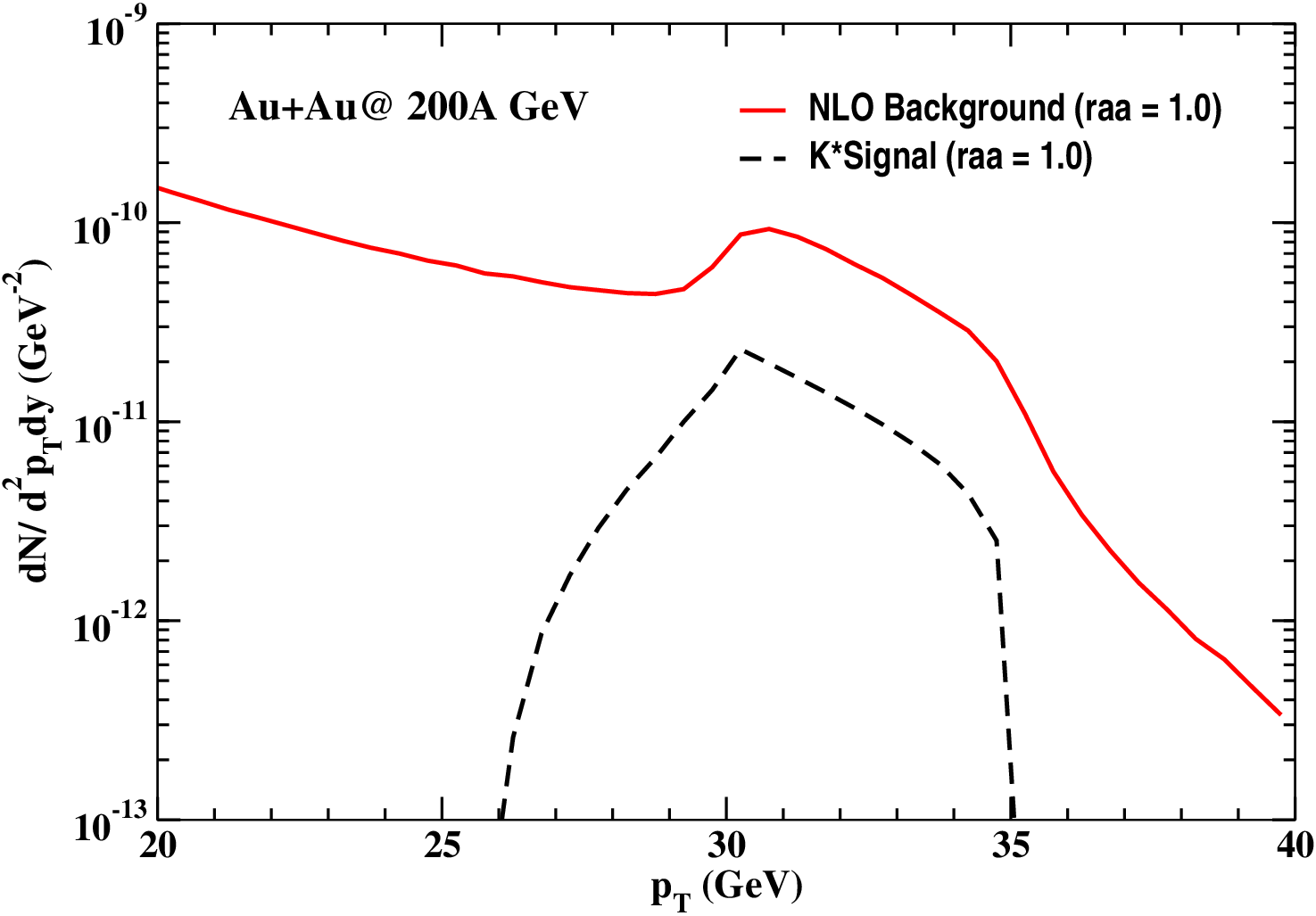}
\caption{(Color online) Upper Panel: Yield $dN/d^2 p_T dy$ of photons opposite of a jet with 
  energy between 30 and 35 GeV in central Au+Au collisions at 
  $\sqrt{s_{NN}}=200$ GeV. We show the sum of prompt hard photons 
  and fragmentation photons (solid and dashed line) and back-scattering 
  photons (solid lines with marks) at LO accuracy for the 
  hard process for different parton and trigger energy loss scenarios.
  Triangles: signal for no parton energy loss and no trigger jet energy loss 
  (raa 1.0). Squares: parton energy loss on, no trigger jet energy loss (raa 1.0).
  Circles: both parton energy loss and trigger jet energy loss at realistic 
  strength (raa 0.7).
  Lower Panel: The same as upper panel, however background 
  (prompt hard + fragmentation) calculated at NLO accuracy for the case 
  raa 1.0 (solid line). Back-scattering photons for raa 1.0 (dashed line) at LO 
  accuracy multiplied by a $K$-factor.}
\label{fig:rhicspec}
\end{center}
\end{figure}

\begin{figure}[tb]
\begin{center}
\includegraphics[width=\columnwidth,clip= true  
]{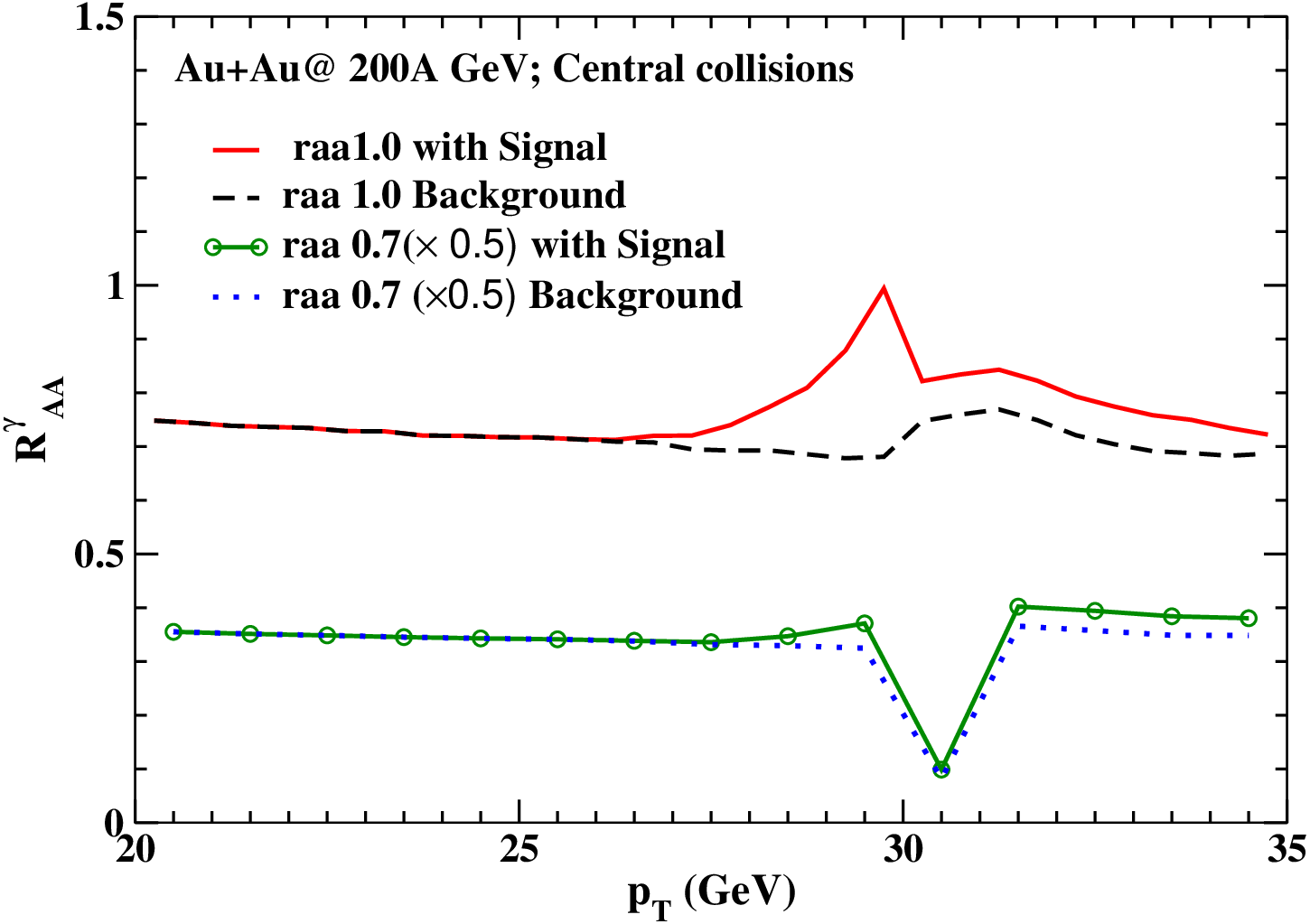}\\[2em]
\includegraphics[width=\columnwidth,clip= true]{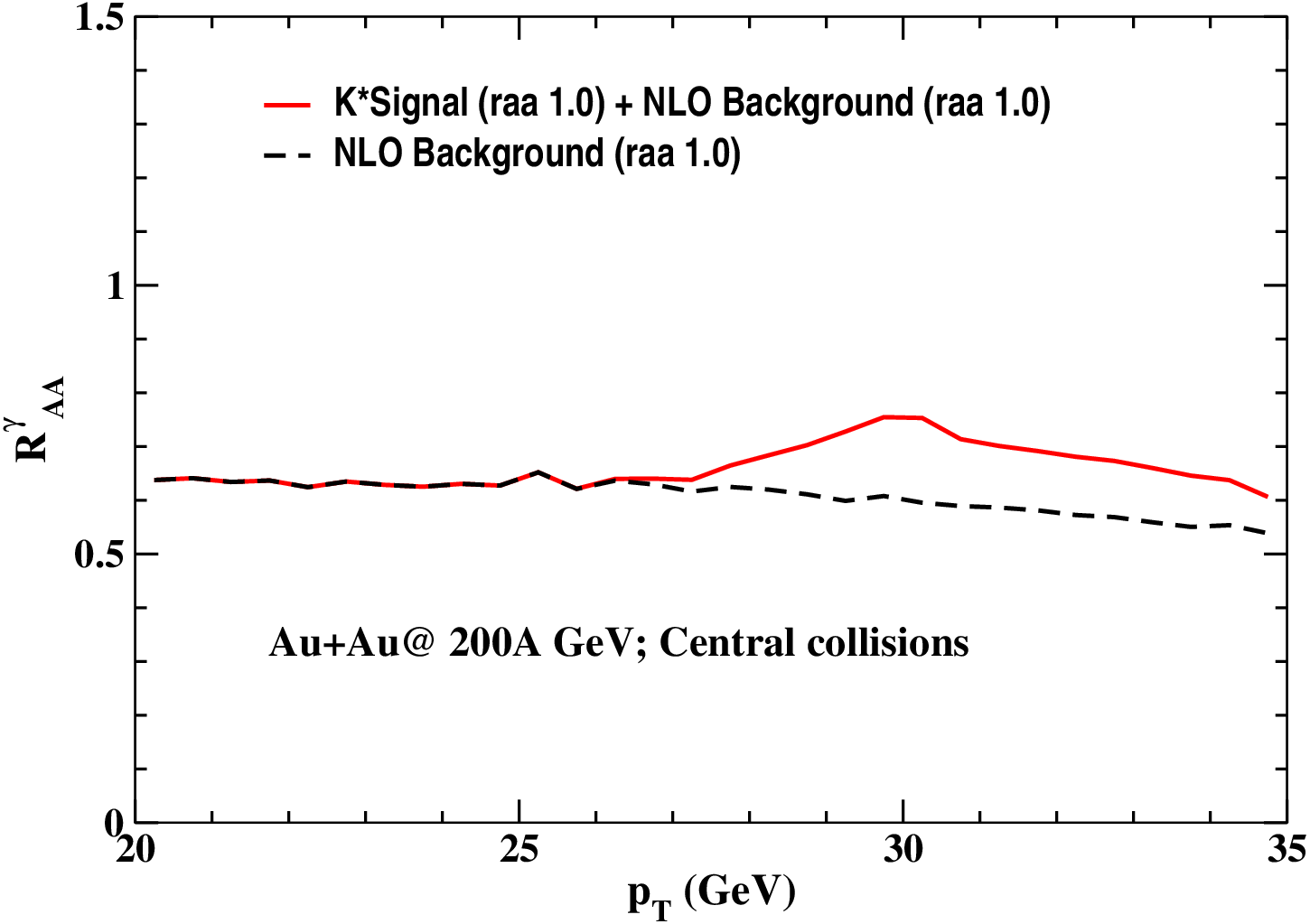}
\caption{(Color online) Upper Panel: 
 Nuclear modification factor $R_{AA}$ calculated from the results
 in Fig.\ \ref{fig:rhicspec} for background and signal at LO accuracy
 for (i) jet raa 1.0  and (ii) jet raa 0.7 
 (scaled by 0.5 for better visibility).
 In both cases (signal+background)/background and the reference 
 background/background are shown.
 Lower Panel: The same as the upper panel but for raa 1.0 at NLO accuracy.
\label{fig:rhicraa}
}
\end{center}
\end{figure}

Fig.\ \ref{fig:rhicspec} shows our results for jet-triggered photon spectra
in central Au+Au collisions at RHIC for scenarios with jet raa 1.0 and 0.7
at leading-order (LO) accuracy.  
At LO and without trigger energy loss (raa 1.0) the prompt hard 
photon kinematics is completely
determined by the trigger jet energy and leads to a well-defined band of
photons between 30 and 35 GeV in transverse momentum. 
Fragmentation photons generally provide a low-level background just below
the trigger window (they correspond to very high-$z$ photons). We use BFG-II
fragmentation function for photons \cite{Bourhis:1997yu}.
The kinematic range of back-scattering photons (the signal) calculated 
under the same assumptions (LO, raa 1.0) and without energy loss of partons, 
coincide with those of prompt hard photons, as expected, although their 
strength is lower by about an order of magnitude.
If parton energy loss is switched on with parameters determined from 
single hadron suppression, the
back-scattering signal develops a shoulder of about 4 GeV width, indicating
that quarks have lost up to 4 GeV of energy before conversion to photons.
This pushes some back-scattering photon strength into the region of
fragmentation photons which makes for a much better signal/background ratio
just below the trigger window.

If jet energy loss is taken into account in addition, with cone radii
currently available (raa 0.7), both the hard prompt photon background 
and the back-scattering photon spectra become slightly
more diffuse and tend to be shifted to higher $p_T$ since a trigger jet measured
between 30-35 GeV might have originated as a jet with larger energy. 
The jet triggered photon spectra thus carry fairly obvious information 
about the energy loss of partons and trigger jets in their broadening around
the trigger window. 

These strong kinematic correlations are washed out by NLO corrections
to the hard process in which another hard parton can be emitted in the 
final state. The effect is estimated in the lower panel of 
Fig.~\ref{fig:rhicspec}
where the background is now calculated at NLO accuracy and raa 1.0.
We also show the back-scattering photons at LO accuracy but with a $K$ factor.
Our calculation of back-scattering photons in its current form is not suitable
to deal with radiative corrections as it is not clear how to treat
medium induced radiation of a collinear pair of quarks that would end up 
in the same jet cone. However our results seem to indicate that the 
decorrelation of the signal with the trigger window that comes from
radiative corrections to the hard process is generally weaker than the
decorrelation that is induced by parton and trigger jet energy loss. This
is even more the case at LHC energies where energy loss is large.
Here the $K$ factor is determined from the ratio 
(background at NLO)/(background at LO) in the fragmentation dominated region 
of the background. We chose to determine $K$ at 20 GeV for RHIC and 
40 GeV for LHC.

We proceed to show the results for the nuclear modification factor
$R_{AA}$. Experimentally, $R_{AA}$ can be determined with smaller systematic
uncertainties compared to spectra, and might thus be a more promising observable.
Fig.~\ref{fig:rhicraa} shows $R_{AA}$ as defined in
Eq.~(\ref{eq:raa}) for central Au+Au collisions at RHIC for (i) raa 1.0 and
(ii) raa 0.7 (scaled by 0.5). For comparison we also show the result one would
obtain if back-scattering photons were absent (i.e.\ the ratio of
fragmentation and prompt hard photons for Au+Au and $p+p$). The difference
between the $R_{AA}$ with and without inclusion of signal,  
is the signature for jet-triggered back-scattering photons.

Let us understand the key features of $R_{AA}$. First, we note that the 
nuclear modification factor of background photons (i.e.\ background photons in A+A vs background
photons in $p+p$) is not around 1. This is because background photons at RHIC
probe hard processes with quarks in the initial wave function. In A+A those
processes are suppressed due to the larger fraction of $d$ valence quarks 
compared to $u$ valence quarks in nuclei and their smaller electric charge.
We notice that trigger jet energy loss can lead to suppression of $R_{AA}$ 
in the trigger window due to the shift of strength of background photons to 
larger energies. In fact the width of such a dip is related to the size 
of the typical jet energy loss. The signal of back-scattering photons on the
other hand creates an enhancement in $R_{AA}$ which is peaked just below the
trigger window. Both the dips in the background and the enhancement due
to the signal are typical effects that will also appear at LHC.
In contrast, radiative effects on the distribution of background photons in 
the NLO calculation generally tends to smear out an enhancement due to the 
signal.  

\begin{figure}[tb]
\begin{center}
\includegraphics[width=\columnwidth, clip=true  
]{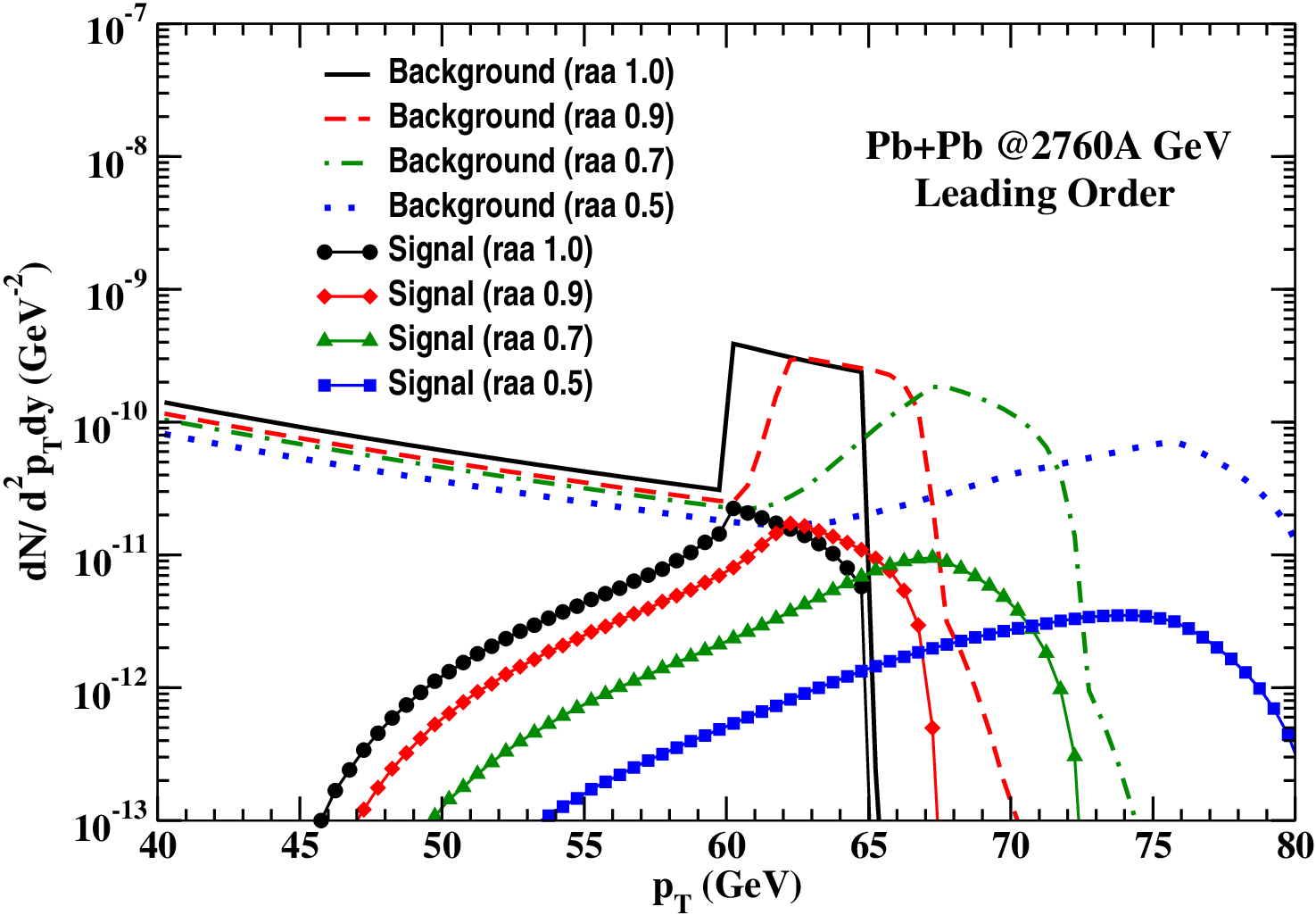}\\[2em]
\includegraphics[width=\columnwidth, clip=true]{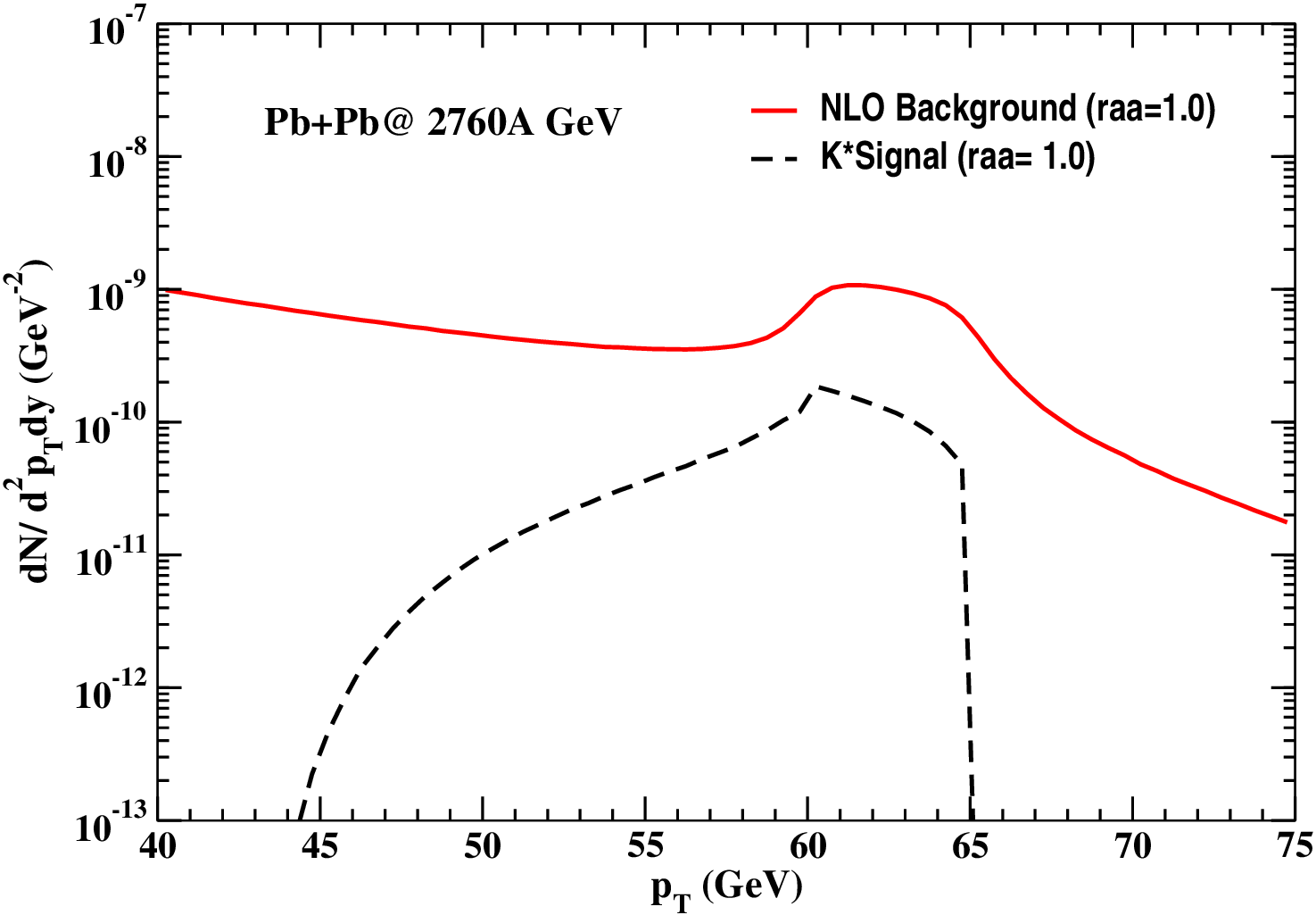}
\caption{(Color online) Upper panel: The same as Fig.\ \ref{fig:rhicspec} for trigger jets
  between 60 and 65 GeV energy in central Pb+Pb collisions at $\sqrt{s_{NN}}
  =$ 2760 GeV for four different trigger jet energy loss scenarios: raa 1.0
  (solid line, circles), raa 0.9 (dashed line, diamonds), raa 0.7 (dash-dotted
  line, triangles), 0.5 (dotted line, squares).
  Both background and back-scattering signal are calculated at LO accuracy 
  for the hard process. All scenarios have parton energy loss taken into 
  account. Lower panel: Background (prompt hard~+~fragmentation) is calculated 
  at NLO accuracy for the case raa 1.0 (solid line) while back-scattering photons are estimated 
  at LO accuracy for raa 1.0 (dashed line), multiplied by a $K$-factor.
  \label{fig:lhcspec}
}
\end{center}
\end{figure}
\begin{figure}[tb]
\begin{center}
\includegraphics[width=\columnwidth, clip=true  
]{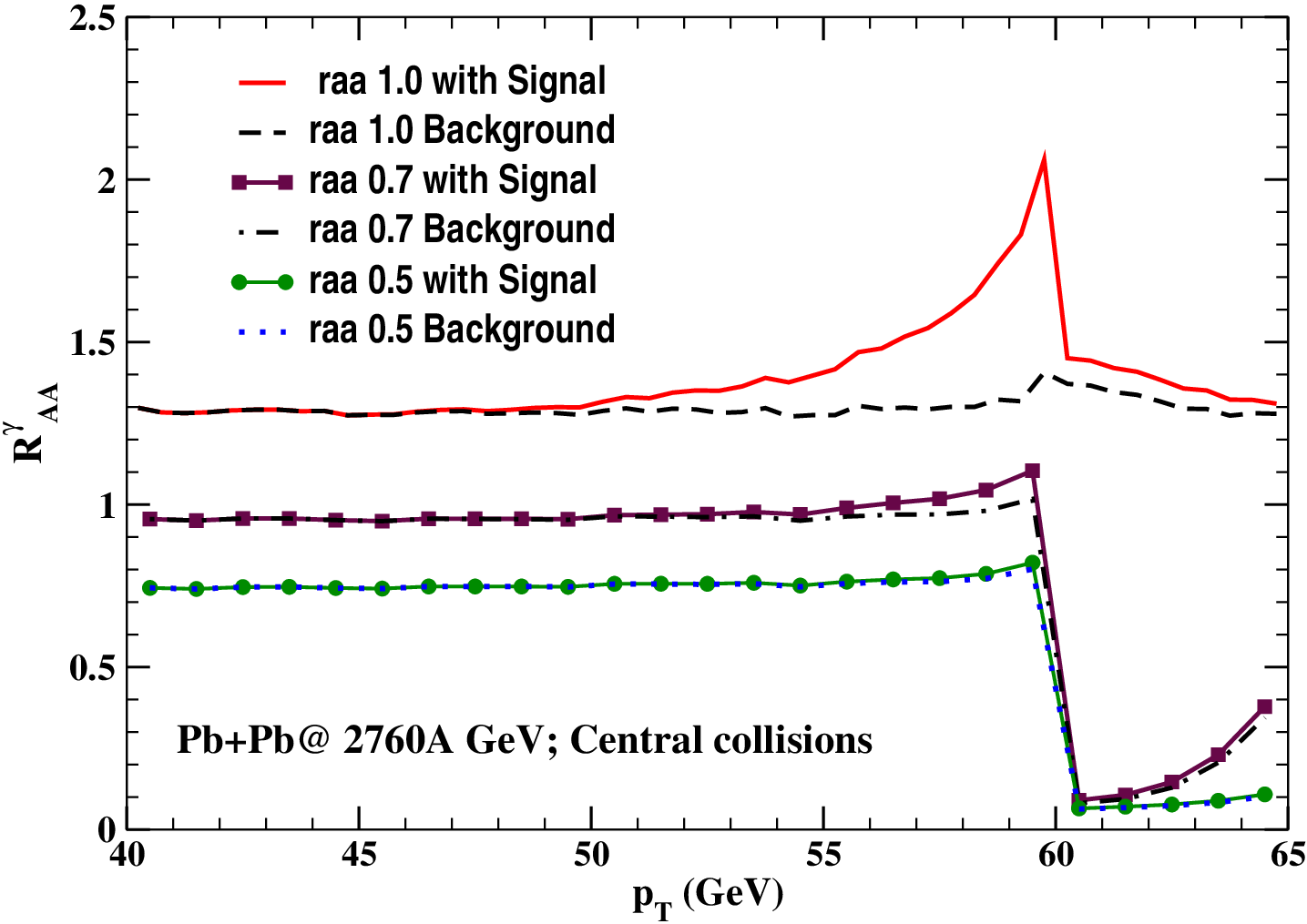}\\[2em]
\includegraphics[width=\columnwidth,clip=true]{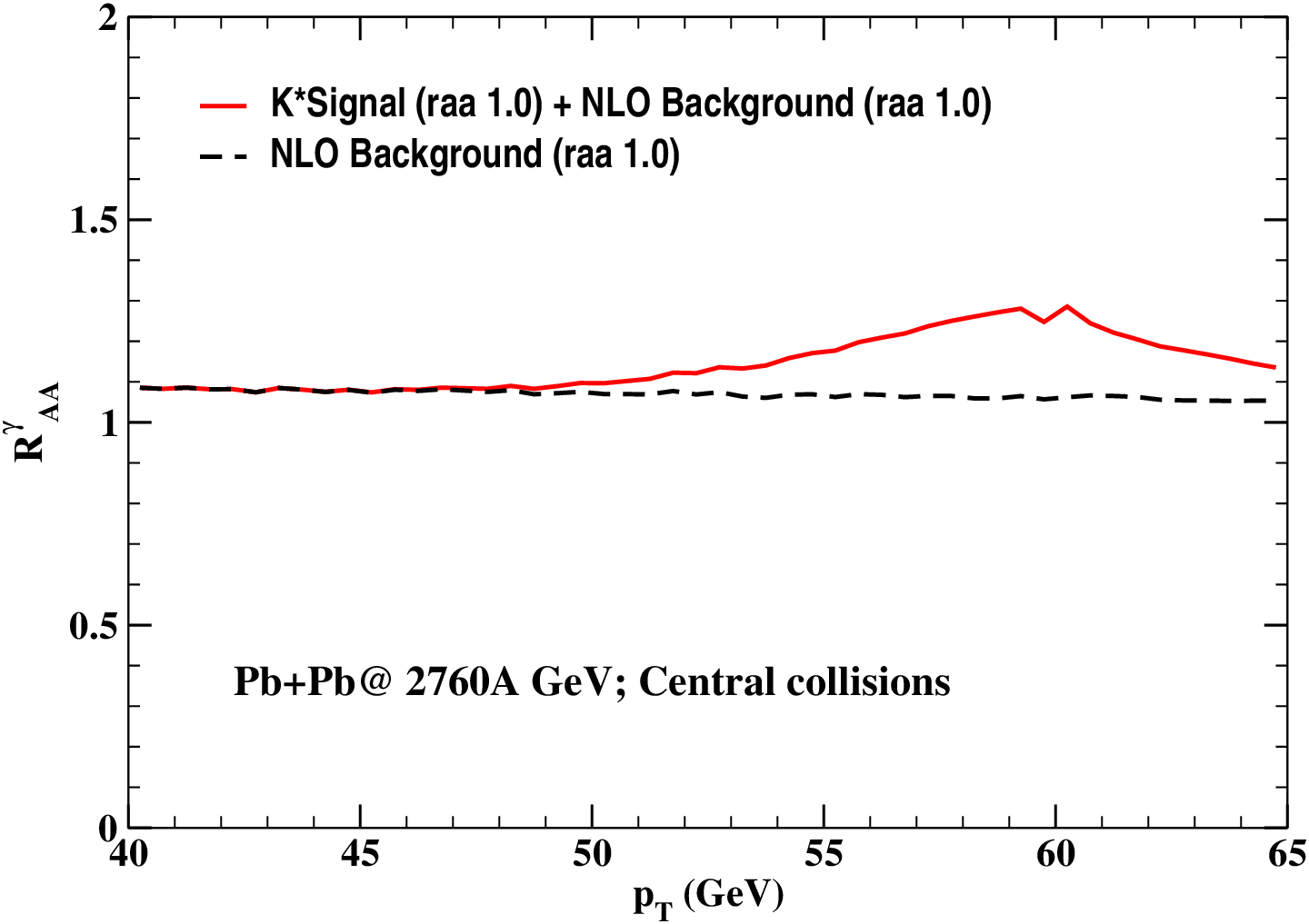}
\caption{(Color online) Upper panel: Nuclear modification factor $R_{AA}$ for central 
  collisions of lead nuclei at LHC at LO accuracy for raa 1.0,
  raa 0.7 and raa 0.5. We again show $R_{AA}$
  with and without the inclusion of back-scattering photons.
  Lower panel: The same calculated for raa 1.0 at NLO accuracy.}
\label{fig:lhcraa}
\end{center}
\end{figure}
Fig.~\ref{fig:lhcspec} shows the jet-triggered photon spectrum for central
Pb+Pb collisions at LHC for the 60--65 GeV trigger window discussed above.
We show both signal and background for the four jet energy
loss scenarios (raa 1.0, 0.9, 0.7 and 0.5) at LO kinematics with parton
energy loss included. All the features discussed for the RHIC case are 
qualitatively present at LHC as well. However, the diffusion of signal strength 
both due to parton energy loss and jet energy loss is much larger than 
at RHIC for the raa 0.7 and 0.5 scenarios, creating shoulders up to 15--20 GeV
wide on both sides of the trigger window. 

Fig.~\ref{fig:lhcraa} shows $R_{AA}$ for trigger energy loss
scenarios raa 1.0, raa 0.7, and raa 0.5. The baseline suppression due to $d$ valence
quarks is not present at LHC where hard processes are dominated by gluon 
fusion. We again find dips in the background $R_{AA}$ in the trigger window
due to trigger energy loss, and enhancement in $R_{AA}$, peaked below the
trigger window, from back-scattering photons. While raa 1.0 shows a rather
promising signature peak the signal for the more realistic raa 0.5 and 
raa 0.7 jet energy loss scenarios are small. 
\section{Summary and Discussion}
We have, for the first time, calculated the correlation of medium-induced
photon radiation from jets with trigger jets. We have focused on 
back-scattering photons from the Compton process. Our numerical studies
indicate that there is a potential signal from back-scattering photons 
in the $R_{AA}$ of photons opposite of trigger jets in high energy nuclear 
collisions. The signal is mostly due
to a downward shift of back-scattering photons in momentum due to parton 
energy loss before the back-scattering occurs. This reduces the background 
from prompt hard photons significantly. However, trigger jet
energy loss and radiative corrections to the underlying hard processes tend
to wash out the correlation. The decorrelation of signal and trigger due 
to jet energy loss dominates over those due to NLO corrections at LHC 
energies. With the currently used small jet cone radii and the typical 
trigger jet $R_{AA}$ measured at RHIC and LHC the signal is visible in our
calculation, but it would be too small to be seen experimentally. 

We should emphasize here that many features of our calculation are designed
to establish a \emph {lower} bound on the signal strength and a more detailed
follow-up calculation could lead to a more promising result. Here are the main
points that establish a lower bound:
(i) The simple equation of state underestimates the temperature and thus the
back-scattering rate. (ii) We omitted induced photon bremsstrahlung,
which will generally increase the signal photon rate below the trigger window.
Obviously back-scattering photons have the advantage of a rather sharp
feature in $R_{AA}$, while additional yield which simply scales up $R_{AA}$ 
in a $p_T$-independent way will be harder to find experimentally. (iii)
Photon fragmentation might happen partially or fully outside of the medium. 
In that case fragmentation photons are subject to energy loss which is
neglected here. This effect will shift the background from fragmentation 
towards smaller $p_T$, effectively decreasing the background. 
We have also not systematically explored different kinematic cuts on the
trigger jet or the photon that could possibly improve the signal over
background ratio. Nevertheless we conclude that single inclusive 
jet $R_{AA}$ of 0.7 or larger in central collisions will likely be necessary 
to carry out this measurement.

Going beyond the back-scattering peak approximation in Eq.\ (\ref{eq:master})
leads to decorrelation of the trigger and back-scattering photon, however, it
will also tend to push some of the signal strength to lower $p_T$, away from 
the prompt hard photon background. The net effect might thus not be simply
a loss of signal due to decorrelation. In principle our proof-of-principle
calculation could be improved in several ways. A full jet shower Monte-Carlo
would remove the need for a leading parton approximation. It could also mimic
NLO kinematics which we have only employed when final state effects leading
to energy loss is absent since no consistent theory is available in that case.

One could consider the use of high-$p_T$ trigger hadrons instead of
trigger jets. They will be subject to the parton energy loss and jet energy
loss which leads to smearing of the back-to-back energy correlation as discussed in the
jet-photon case. In addition, there is smearing due to the fragmentation 
of the hadron which by itself already almost completely destroys the 
correlation in energy with the photon on the other side, see e.g.\ 
\cite{Qin:2008rd}. Therefore hadron triggered photons have not been considered
here.

SD is grateful to the Cyclotron Institute at Texas A\&M University for their
hospitality during his stay.
This project was supported by the U.S. National Science Foundation through 
CAREER grant PHY-0847538, and by the JET Collaboration and DOE grant
DE-FG02-10ER41682. SD acknowledges the financial support of DAE, India during 
the course of this work. 

\end{document}